\documentclass[twoside,12pt]{article}
\usepackage{epsf,epsfig,amssymb,amsmath,graphicx,float}
\usepackage{color}

\newcommand{\gsim}{\raisebox{-0.13cm}{~\shortstack{$>$ \\[-0.07cm] $\sim$}}~}

\begin{document}
\renewcommand{\thefootnote}{\fnsymbol{footnote}}

\begin{titlepage}

\begin{center}

\vspace{1cm}

{\Large {\bf Asymmetric Dark Matter abundance including non-thermal
    production  }}

\vspace{1cm}

{\bf Maqingshan, Hoernisa Iminniyaz\footnote{Corresponding 
author, wrns@xju.edu.cn}}

\vskip 0.15in
{\it
{School of Physics Science and Technology, Xinjiang University, \\
Urumqi 830046, China} \\
}

\abstract{ We investigate the relic abundance of asymmetric Dark Matter where
  the asymmetric Dark Matter is non--thermally produced from the decay of
  heavier particles in addition to the usual thermal production. We discuss 
  the relic density of asymmetric Dark Matter including the decay of heavy 
  particles in low temperature scenario. Here we still
  assume the universe is radiation dominated and there is asymmetry
  before the decay of heavy particles. We obtained
  increased abundances of asymmetric Dark Matter when there is additional
  contribution from the decay of heavier particles. Finally we find the 
  constraints on the asymmetry factor and annihilation cross section using 
  the Planck data.  }
\end{center}
\end{titlepage}
\setcounter{footnote}{0}

\section{Introduction}

Dark Matter problem is undoubtedly one of the most challenging
puzzles for cosmologists and particle physicists in recent years. 
Although we have most striking evidences for Dark Matter, the constitute of 
Dark Matter is still a mistery for us. Asymmetric Dark Matter is
one alternatives which arises from the fact that its 
abundance is just 5 times larger than the baryon asymmetry \cite{adm-models}. 
There maybe an indication of common origin which is responsible for the 
baryonic asymmetry and Dark Matter. In asymmetric Dark Matter scheme, Dark 
Matter particle has its distinct antiparticle. 

In standard cosmological scenario, it is supposed 
asymmetric Dark Matter were produced thermally. Based on the assumption that 
Weakly Interacting Massive Particles (WIMPs) constitute 
Dark Matter, the relic density of asymmetric Dark Matter is 
determined by the freeze-out condition \cite{GSV,Iminniyaz:2011yp}. In this 
scenario, it is assumed the reheating temperature $T_R$ is much 
larger than the freeze out temperature $T_F$ and the asymmetric Dark Matter 
particles and anti--particles were in thermal equilibrium in the early 
universe. The final relic density of asymmetric Dark Matter mainly depends on 
the particle--antiparticle asymmetry and the indirect detection signals from 
annihilation is suppressed due to the reason that very little anti--particle 
is here in the end. 

On the other hand, asymmetric Dark Matter can be produced non-thermally as 
heavy particles decay (e.g. moduli decay) into asymmetric Dark Matter. In 
ref.\cite{Dhuria:2015xua}, the authors discussed a cogenesis
mechanism in which the baryon asymmetry of the universe and Dark Matter 
abundance are simultaneously produced at low reheating temperature. 
Refs.\cite{Blinov:2014nla} explored the non--thermal production of Dark Matter 
including asymmetric Dark Matter in the scheme of Minimal Supersymmetric 
Standard Model and string motivated models. Relic abundance of Dark Matter 
was calculated in ref.\cite{Drees:2006vh} including the decays of unstable 
heavy particle to Dark Matter particles. In our work, we investigate the 
relic density of asymmetric Dark Matter in low reheating temperature scenario 
when the heavy unstable particles 
$\phi$ decay into asymmetric Dark Matter particles and anti--particles.  
We suppose that the asymmetry is already existed before the decay of heavy
particles into asymmetric Dark Matter and they decay into 
particle and anti--particle in the same amount. We still assume the 
universe is radiation dominated and the additional entropy production from 
the decay of heavy particle is negligible. The decay of heavier particles to 
asymmetric Dark Matter changes the total relic density of asymmetric Dark 
Matter. The relic abundances of 
asymmetric Dark Matter particle and anti--particle are both increased when 
there is additional contribution from the decay of heavier particles.  

The outline of the paper is as following. In section 2, we discuss the 
relic density calculation of asymmetric Dark Matter including the decay of heavy
particles into asymmetric Dark Matter in low temperature scenario. We 
find the approximate analytic solution for the relic density of asymmetric 
Dark Matter when there is thermal and non--thermal 
production. In section 3, the constraints on parameter spaces are obtained
by using the Planck data. The last section is devoted to the brief summary 
and conclusions.

\section{Relic abundance of asymmetric Dark Matter including non-thermal 
production}

We discuss the scenario where the unstable heavy particles $\phi$ decay into
asymmetric Dark Matter particles and anti--particles. Here we assume the 
heavy particle $\phi$ 
decays out of thermal equilibrium, therefore, $\phi$ 
production is negligible. However, we include  the thermal and non--thermal 
production of asymmetric Dark Matter particles.  

In the standard computation of relic density for the asymmetric 
Dark Matter, it was assumed the reheating temperature of the universe 
$T_R$ is much higher than the freeze out temperature. In this case, the 
reheating era has no effect on the final relic density of asymmetric Dark 
Matter. On the other hand, the constraints on the reheating temperature 
comes from Big Bang Nucleosynthesis and $T_R \gsim 1$ MeV 
\cite{Giudice:2000ex,Hannestad:2004px,Ichikawa:2005vw}. Therefore, 
we consider the case where $T_R < T_F$ 
\cite{Fornengo:2002db,Bastero-Gil:2000ywn,Kudo:2001ie,Gelmini:2006pw}. The 
asymmetric Dark 
Matter particles and anti--particles never reach thermal equilibrium due to 
the low reheating temperature, we treat the asymmetric Dark Matter abundances 
at some initial temperature $T_0$ as a free parameter.

The evolution of relic abundance including the decay of heavy particles is 
more complicated than the usual thermal production case. Here we still assume 
the universe is radiation dominated and the entropy production is negligible. 
We suppose that asymmetry is existed before the decay of heavy particles and
they decay into particle and anti--particle in the same 
amount. The coupled Boltzmann equations for the number densities $n_{\chi}$ and 
$n_{\bar\chi}$ of asymmetric Dark Matter particles $\chi$ and anti--particles 
$\bar\chi$ are as following,
\begin{equation}\label{eq:Boltzmann-n1}
      \frac{dn_{\chi}}{dt} + 3Hn_{\chi}  =  -\langle \sigma v \rangle (n_{\chi}n_{\bar\chi} - n_{\chi,{\rm eq}}n_{\bar\chi,{\rm eq}}) +
          N \Gamma_{\phi} n_{\phi}, 
\end{equation}
\begin{equation}\label{eq:Boltzmann-n2}
      \frac{dn_{\bar\chi}}{dt} + 3Hn_{\bar\chi}  =  -\langle \sigma v \rangle (n_{\chi}n_{\bar\chi} - n_{\chi,{\rm eq}}n_{\bar\chi,{\rm eq}}) +
       N \Gamma_{\phi} n_{\phi}, 
\end{equation}
\begin{equation}\label{eq:Boltzmann-n3}
     \frac{dn_{\phi}}{dt} + 3Hn_{\phi}  =  - \Gamma_{\phi}\, n_{\phi},
\end{equation}
where $H$ is the expansion rate of the universe, and 
$\langle \sigma v \rangle$ is the thermal average of the cross section 
multiplied with the relative velocity of 
annihilating  asymmetric Dark Matter particles and anti--particles. Here we 
assume only $\chi\bar\chi$ pairs can annihilate into Standard 
Model particles. $n_{\chi,{\rm eq}}$ is the equilibrium
number density. $N$ is the average number of asymmetric Dark Matter particle
and anti--particle produced in $\phi$ decay, $\Gamma_{\phi}$ is the decay 
rate.  Here it is assumed that $\phi$ does not dominate the total 
energy density. Therefore, the comoving entropy density almost stays constant 
throughout. The analytic solution for Eq.(\ref{eq:Boltzmann-n3}) is easily 
obtained. Then Eqs.(\ref{eq:Boltzmann-n1}), (\ref{eq:Boltzmann-n2}) can be 
solved approximately in low temperature scenarios.

In order to simplify 
Eqs.(\ref{eq:Boltzmann-n1}),(\ref{eq:Boltzmann-n2}),(\ref{eq:Boltzmann-n3}), 
we introduce the following dimensionless variables 
$Y_{\chi,\bar\chi} = n_{\chi,\bar\chi}/s$, $Y_{\phi} = n_{\phi}/s$ and 
$x = m/T$ with $s$ being the entropy density, then
\begin{equation}\label{eq:Boltzmann-Yt1}
\frac{d (s Y_{\chi})}{dt} + 3 H sY_{\chi} = - \langle \sigma v \rangle s^2
      (Y_{\chi} Y_{\bar\chi} - Y_{\chi,{\rm eq}} Y_{\bar\chi,{\rm eq}} ) +
       N \Gamma_{\phi }\, s Y_{\phi}, 
\end{equation}
\begin{equation}\label{eq:Boltzmann-Yt2}
\frac{d (s Y_{\bar\chi})}{dt} + 3 H sY_{\bar\chi} = - \langle \sigma v \rangle s^2
      (Y_{\chi} Y_{\bar\chi} - Y_{\chi,{\rm eq}} Y_{\bar\chi,{\rm eq}} ) +
       N \Gamma_{\phi }\, s Y_{\phi}, 
\end{equation}
\begin{eqnarray}\label{eq:phi}
      \frac{d (s Y_{\phi})}{dt} + 3 H s Y_{\phi} = 
      - \Gamma_{\phi }\, s Y_{\phi}\,,
\end{eqnarray}
Assuming entropy per comoving volume is conserved, we obtain
\begin{equation}\label{sH}
      \dot{s} = - 3s H ,
\end{equation}
here we used $H = \dot{R}/R$, where $R$ is the scale factor of the universe. 
Inserting $s = (2 \pi^2/45) g_* T^3$ to Eq.(\ref{sH}), here $g_*$ is the
effective number of relativistic degrees of freedom, we obtain 
\begin{equation}
     \frac{1}{T}\frac{dT}{dt} + \frac{1}{3 g_*} \frac{dg_*}{dt} + H = 0,
\end{equation}
Using $T = m/x$
\begin{equation}
      \frac{dx}{dt} = \frac{Hx}{ 1 - \frac{x}{3g_*} \frac{dg_*}{dx}}\,.
\end{equation}
Then Eqs.(\ref{eq:Boltzmann-Yt1}), (\ref{eq:Boltzmann-Yt2}) and 
(\ref{eq:phi}) can be written as   
\begin{equation}\label{eq:Boltzmann-Yxchi}
      \frac{dY_{\chi}}{dx} = -\frac{\langle \sigma v \rangle s}{Hx}
            (Y_{\chi} Y_{\bar\chi} - Y_{\chi,{\rm eq}} Y_{\bar\chi,{\rm eq}} )
           +\frac{N \Gamma_{\phi }}{Hx} Y_{\phi}\,,
\end{equation}
\begin{equation}\label{eq:Boltzmann-Yxchibar}
      \frac{d Y_{\bar\chi}}{dx} = -\frac{\langle \sigma v \rangle s}{Hx}
            (Y_{\chi} Y_{\bar\chi} - Y_{\chi,{\rm eq}} Y_{\bar\chi,{\rm eq}} )
           +\frac{N \Gamma_{\phi }}{Hx} Y_{\phi}.
\end{equation}
\begin{equation}\label{eq:phix}
      \frac{dY_{\phi}}{dx} = -\frac{\Gamma_{\phi}}{Hx} Y_{\phi},
\end{equation}
here we assume $dg_*/dx = 0$. Eq.(\ref{eq:phix}) can be solved easily,
\begin{equation}
     \int^{Y_{\phi}(x)}_{Y_{\phi}(x_0)} \frac{dY_{\phi}}{Y_{\phi}} = 
        - \int^x_{x_0}\frac{\Gamma_{\phi} dx}{Hx},
\end{equation}
\begin{equation}
      Y_{\phi}(x) = Y_{\phi}(x_0) e^{-\int^x_{x_0} \frac{\Gamma_{\phi}dx}{Hx}}\,,
\end{equation}
where $x_0$ is the inverse scaled initial temperature which is close to the
freeze out temperature. 
Then equations (\ref{eq:Boltzmann-Yxchi}) and (\ref{eq:Boltzmann-Yxchibar}) 
become
\begin{equation}\label{eq:Ychi-Yphi}
      \frac{d Y_{\chi}}{dx} = -\frac{\langle \sigma v \rangle s}{Hx}
            (Y_{\chi} Y_{\bar\chi} - Y_{\chi,{\rm eq}} Y_{\bar\chi,{\rm eq}} )
           +\frac{N \Gamma_{\phi}}{Hx} 
           Y_{\phi}(x_0) e^{-\int^x_{x_0} \frac{\Gamma_{\phi}}{Hx}\,dx}\,,
\end{equation}
\begin{equation}\label{eq:Ychibar-Yphi}
      \frac{d Y_{\bar\chi}}{dx} = -\frac{\langle \sigma v \rangle s}{Hx}
            (Y_{\chi} Y_{\bar\chi} - Y_{\chi,{\rm eq}} Y_{\bar\chi,{\rm eq}} )
           +\frac{N \Gamma_{\phi}}{Hx} 
           Y_{\phi}(x_0) e^{-\int^x_{x_0} \frac{\Gamma_{\phi}}{Hx}\,dx}\,.
\end{equation}
In the radiation dominated era, 
$ H =  \pi m^2_{\chi}/(M_{\rm Pl}\,x^2)\sqrt{g_*/90}$, where $m_{\chi}$ is the mass
of asymmetric Dark Matter particle and  
$M_{\rm Pl} = 2.4 \times 10^{18}$ GeV is the reduced Planck mass. Now we 
further simplify Eqs.(\ref{eq:Ychi-Yphi}) and (\ref{eq:Ychibar-Yphi}), 
subtracting equation (\ref{eq:Ychibar-Yphi}) from 
equation (\ref{eq:Ychi-Yphi}), then
\begin{equation}\label{eq:eta}
      \frac{d (Y_{\chi} - Y_{\bar\chi})}{dx} = 0.
\end{equation}
This requires
\begin{equation}\label{eq:eta}
      Y_{\chi} - Y_{\bar\chi} = \eta.
\end{equation}
Where $\eta$ is a constant. We can express equations 
(\ref{eq:Ychi-Yphi}) and (\ref{eq:Ychibar-Yphi}) in terms of Eq.(\ref{eq:eta}),
\begin{equation}\label{eq:Ychifin}
      \frac{d Y_{\chi}}{dx} = -\frac{\lambda_1 \langle \sigma v \rangle}{x^2}
    \left(Y^2_{\chi}  - \eta Y_{\chi} - \lambda_2 x^3 e^{-2x}
          \right) +\lambda_3 N  Y_{\phi}(x_0) x\, 
          {\rm exp}\left[ \frac{\lambda_3}{2}(x_0^2-x^2 ) \right],
\end{equation}
\begin{equation}\label{eq:Ychibarfin}
      \frac{d Y_{\bar\chi}}{dx} = 
      -\frac{\lambda_1 \langle \sigma v \rangle}{x^2}
      \left(Y^2_{\bar\chi}  + \eta Y_{\bar\chi} - \lambda_2 x^3 e^{-2x}
          \right)
           + \lambda_3 N  Y_{\phi}(x_0) x\, 
          {\rm exp}\left[ \frac{\lambda_3}{2}(x_0^2-x^2)  \right],
\end{equation}
where $\lambda_1 = \frac{4 \pi}{\sqrt{90}} m_{\chi} M_{\rm Pl} \sqrt{g_*}\,$, 
$\lambda_2 = (0.145 g_\chi /g_* )^2$, and 
$\lambda_3 =    M_{\rm Pl} \Gamma_{\phi }/(\pi m^2_{\chi})\sqrt{90/g_*}$. Here 
we used the equilibrium abundance 
$Y_{\chi,\bar\chi,{\rm eq}} = 0.145 g_{\chi}/g_* x^{3/2} {\rm e}^{-x(1 \pm \mu_{\chi}/m)}\,$, 
where $\mu_{\chi}$ is the chemical potential for particle, $g_{\chi}$ is
number of internal degrees of freedom. 
Usually the thermal average of $\sigma v$ is approximated by a nonrelativistic 
expansion:
\begin{equation} \label{eq:cross}
   \langle \sigma v \rangle = a + 6\,b x^{-1} + {\cal O}(x^{-2})\, ,
\end{equation}
where $a$ is the $s$--wave contribution for the 
limit $v\to 0$ and $b$ is 
$p$--wave contribution for the suppressed $s$--wave annihilation.
%
%
\begin{figure}[h]
  \begin{center}
     \hspace*{-0.5cm} \includegraphics*[width=8cm]{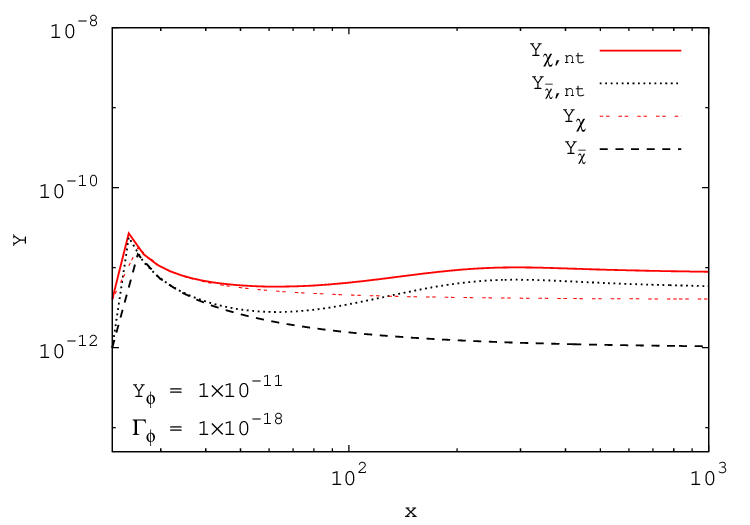}
    \put(-115,-12){(a)}
    \hspace*{-0.5cm} \includegraphics*[width=8cm]{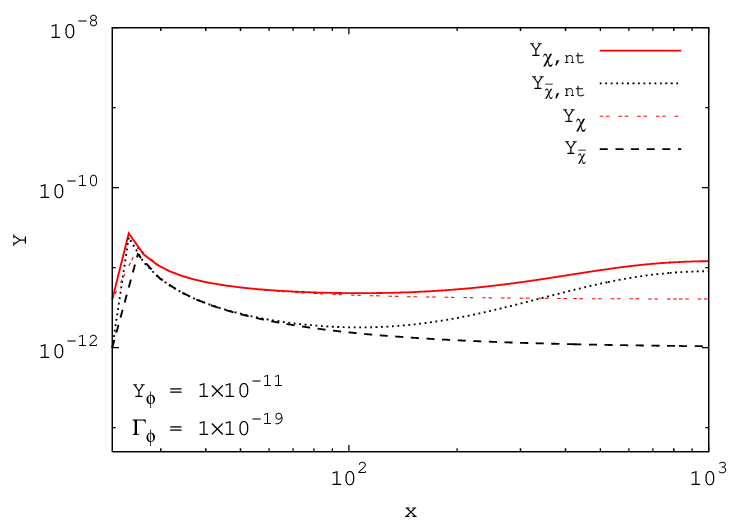}
    \put(-115,-12){(b)}
     \vspace{0.5cm}
     \hspace*{-0.5cm} \includegraphics*[width=8cm]{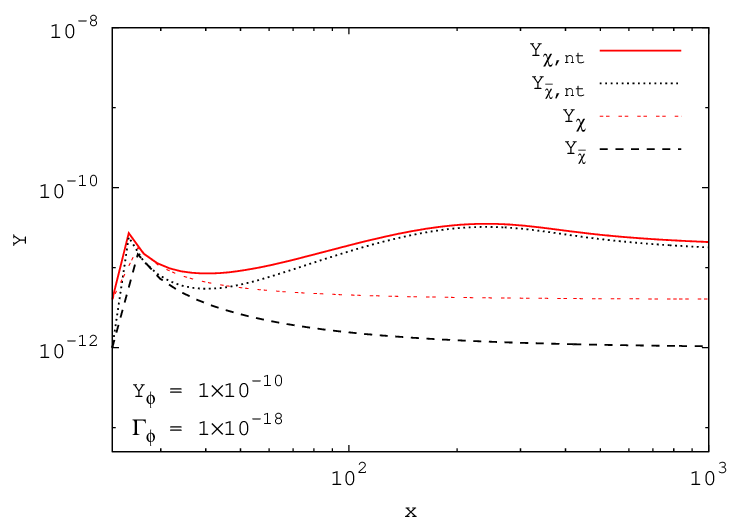}
    \put(-115,-12){(c)}
    \hspace*{-0.5cm} \includegraphics*[width=8cm]{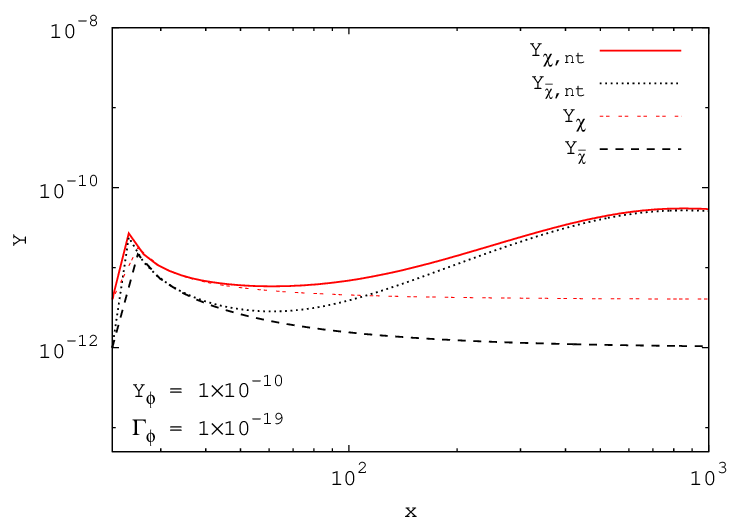}
    \put(-115,-12){(d)}
     \caption{\label{fig:a} \footnotesize
  Evolution of $Y_{\chi}$ and $Y_{\bar\chi}$ as a function of $x$
  for $Y_{\bar\chi}(x_0 = 22) = 1 \times 10^{-12}$, $N = 1$ and $b = 0$
   $ a = 4 \times 10^{-26}$ ${\rm cm}^3 \,{\rm s}^{-1}$, 
    $ \eta = 3 \times 10^{-12}$, $m_{\chi} = 100$ GeV, 
   $x_0 = 22 $, $g_{\chi} = 2$, $g_* = 90$. Here nt means non--thermal
  production. }  
      \end{center}
\end{figure}

Fig.\ref{fig:a} depicts the evolution of relic abundances $Y_{\chi,\bar{\chi}}$
as a function of $x$ and it is based on the numerical solutions of 
Eqs.(\ref{eq:Ychifin}) and (\ref{eq:Ychibarfin}). Here solid red curve is for 
asymmetric Dark Matter particle abundance $Y_{\chi,{\rm nt}}$ with the decay
of heavy particles and the black dotted curve is for antiparticle 
abundance $Y_{\bar\chi,{\rm nt}}$. The red 
dot--dashed and black dashed curves are for asymmetric Dark Matter particle 
and anti--particle relic abundances $Y_{\chi}$ and $Y_{\bar\chi}$ without 
including the decay of heavy particles. 

Fig.\ref{fig:a} shows the abundances of asymmetric Dark Matter particle and 
anti--particle are increased because of the contribution from 
the decay of metastable heavier particles to asymmetric Dark Matter. The
increase of the anti--particle abundance is more remarkable. We start from 
the initial value of $Y_{\bar\chi}(x_0)$ at the 
initial temperature $x_0$. We assume asymmetric Dark Matter particles and 
anti--particles never reach thermal equilibrium because of the low reheating 
temperature. The particle and anti--particle abundances are increased quickly 
shortly after starting and then decreased both for thermal and non--thermal
production. When $x$ increases, the abundances finally goes to constant value. 
The increases for particle and anti--particle abundances are slightly larger 
in panels $(b)$ and $(d)$ of Fig.\ref{fig:a} for smaller decay rate with
respect to panels $(a)$ and $(c)$.   
Comparing frames $(a)$ with $(c)$, and $(b)$ with $(d)$, we can see for lager 
initial value of $Y_{\phi}(x_0)$, there is larger increases of the particle and 
anti--particle abundances. In frames $(c)$
and $(d)$, the anti--particle abundance almost catches the particle abundance
finally. This opens the possibility to detect the asymmetric Dark Matter 
indirectly.

\begin{figure}[h]
  \begin{center}
    \hspace*{-0.5cm} \includegraphics*[width=8.7cm]{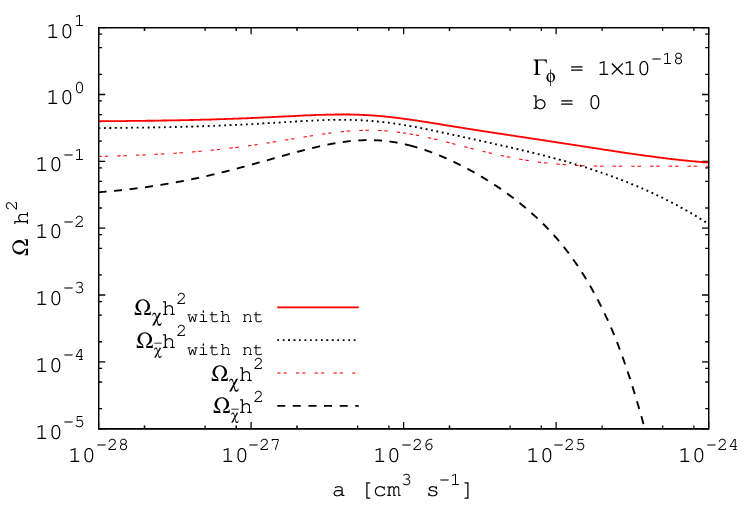}
     \caption{\label{fig:b} \footnotesize
  Relic density $\Omega h^2$ for particle $\chi$ and anti--particle $\bar\chi$ 
  as a function of the $s-$wave cross section. 
  Here $Y_{\bar\chi}(x_0 = 22) = 1 \times 10^{-12}$, $N = 1$ and 
  $Y_{\phi} = 10^{-11}$, $\Gamma_{\phi} = 10^{-18}$, $ \eta = 3 \times 10^{-12}$, 
  $m_{\chi} = 100$GeV, $x_0 = 22 $, $g_{\chi} = 2$, 
  $g_* = 90$.  }  
     \end{center}
\end{figure}
The changes of relic density $\Omega h^2$ as a function of the cross section 
is shown in Fig.\ref{fig:b}. It is for $s$--wave
annihilation cross section with $b = 0$ when asymmetry factor 
$\eta = 3\times 10^{-12}$. Solid red curve and 
dotted black curve are for the relic densities of asymmetric Dark Matter 
particle and anti--particle with the decay of heavy particles. Double dotted 
red curve and black dashed curve are for the particle and anti--particle only 
with thermal production.
Fig.\ref{fig:b} shows when the cross sections increases, the depletion of 
anti--particle abundance is slower than the case of without including heavy
particle decay.

Following we try to find the approximate analytical solutions of
Eqs.(\ref{eq:Ychifin}) and (\ref{eq:Ychibarfin}). 
Eq.(\ref{eq:Ychibarfin}) can be solved first. 
At early times $(x < x_0)$, $Y_{\bar\chi,{\rm eq}} \gg Y_{\bar\chi}$, because
usually $\eta$ is around $10^{-12}$ \cite{Iminniyaz:2011yp},
$\eta Y_{\bar\chi}$ is small as well, therefore equation
(\ref{eq:Ychibarfin}) becomes 
\begin{equation}\label{eq:Boltzmann-Y-x01}
      \frac{d Y_{\bar\chi}}{dx} = 
   \lambda_1 \lambda_2 \langle \sigma v \rangle  x e^{-2x}
           +\lambda_3 N  Y_{\phi}(x_0) x\, 
          {\rm exp}\left( \frac{\lambda_3}{2}(x_0^2-x^2 ) \right)\,.
\end{equation}
Integrating above equation, we obtain
\begin{eqnarray}\label{eq:Boltzmann-Y-x02}
      Y_{\bar\chi}(x)  & = & {Y_{\bar\chi}(x_0)} +  
      \lambda_1  \lambda_2 \left[ \frac{a}{2}(x_0\,e^{-2x_0} - x\,e^{-2x}) + 
     (\frac{a}{4} + 3b)(e^{-2x_0} - e^{-2x})     \right] \nonumber \\
   & + & N Y_{\phi}(x_0)\, 
    \left[ 1 - {\rm exp}\left( \frac{\lambda_3}{2} (x_0^2 - x^2)
     \right)  \right].
\end{eqnarray}
The abundance for $Y_{\chi}$ is obtained using Eq.(\ref{eq:eta}),
\begin{eqnarray}\label{eq:Boltzmann-Ychi-x0}
      Y_{\chi}(x) & = & \eta + {Y_{\bar\chi}(x_0)} +  
     \lambda_1  \lambda_2 \left[ \frac{a}{2}(x_0\,e^{-2x_0} - x\,e^{-2x}) + 
     (\frac{a}{4} + 3b)(e^{-2x_0} - e^{-2x})     \right] \nonumber \\
     & + & N  Y_{\phi}(x_0)\, 
     \left[ 1- {\rm exp}\left( \frac{\lambda_3}{2} (x_0^2 - x^2)
     \right)  \right].
\end{eqnarray}
While the solution is valid only at early times, the relic abundances of 
particles and anti--particles are approximated for $x \gg x_0$ as, 
\begin{eqnarray}\label{eq:Boltzmann-Y-x02-f}
      Y_{\bar\chi,\infty} \equiv Y_{\bar\chi}(x \gg x_0)  =  {Y_{\bar\chi}(x_0)} +  
      \lambda_1  \lambda_2 \left[ \frac{a}{2}x_0\,e^{-2x_0}  + 
     (\frac{a}{4} + 3b)e^{-2x_0}      \right]  +  N  Y_{\phi}(x_0).
\end{eqnarray}
\begin{eqnarray}\label{eq:Boltzmann-Ychi-x0-f}
      Y_{\chi,\infty} \equiv Y_{\chi}(x \gg x_0)  =  \eta + {Y_{\bar\chi}(x_0)} +  
     \lambda_1  \lambda_2 \left[ \frac{a}{2}x_0\,e^{-2x_0}  + 
     (\frac{a}{4} + 3b)e^{-2x_0}      \right] 
     + N Y_{\phi}(x_0).
\end{eqnarray}
The relative prediction for the present relic density is given by
\begin{equation}
      \Omega_{\rm DM} h^2  =  \frac{\rho}{\rho_c} =
      2.76 \times 10^8 m_{\chi} [Y_{\chi,\infty} + Y_{\bar\chi,\infty}
      ]\,{\rm GeV}^{-1},
\end{equation}
where the scaled Hubble constant $h$ in units of 
$100$ km s$^{-1}$ Mpc$^{-1}$ is $ 0.673 \pm 0.098 $. Here 
$\rho=n m = s_0 Y m$ and the
critic density is $\rho_c = 3 H^2_0 M^2_{\rm Pl}$, 
where $s_0 \simeq 2900$ cm$^{-3}$ is the present entropy density and $H_0$ is
the Hubble constant.

In the late universe, for $x \gg x_0 $,
\begin{equation}
      Y_{\bar\chi} \gg Y_{\bar\chi,{\rm eq}}.
\end{equation}
Again the third and fourth terms in Eq.(\ref{eq:Ychibarfin}) decreases 
exponentially as $x$ increases, therefore, equation (\ref{eq:Ychibarfin}) 
becomes
\begin{equation}\label{eq:Ybar-late}
      \frac{d Y_{\bar\chi}}{dx} = 
      -\frac{\lambda_1 \langle \sigma v \rangle}{x^2}
      \left(Y^2_{\bar\chi}  + \eta Y_{\bar\chi} 
          \right).
\end{equation}
Integrating above equation, we obtain
\begin{equation}\label{eq:Ychibar-late}
      Y_{\bar\chi}(x) = 
\frac{\eta}{\left[1 + \eta/Y_{\bar\chi}(x_0)\right]\,
  e^{\int^x_{x_0}\frac{\eta \lambda_1  \langle \sigma v \rangle}{x^2}dx} - 1}\,.
\end{equation}
In the same way as Eq.(\ref{eq:Boltzmann-Ychi-x0}), we have 
\begin{equation}\label{eq:Ychi-late}
      Y_{\chi}(x) = 
  \frac{\eta}{ 1 - \left\{ Y_{\bar\chi}(x_0)/[\eta + Y_{\bar\chi}(x_0)] \right\}\,
  e^{-\int^x_{x_0}\frac{\eta \lambda_1  \langle \sigma v \rangle}{x^2}dx} }\,.
\end{equation}
Inserting Eq.(\ref{eq:cross}) into Eqs.(\ref{eq:Ychibar-late}) and 
(\ref{eq:Ychi-late}), we obtain
\begin{equation}\label{eq:Ychibar-late1}
      Y_{\bar\chi}(x) = 
\frac{\eta}{\left[1 + \eta/Y_{\bar\chi}(x_0)\right]\,
  {\rm exp}\left\{ \eta \lambda_1
  \left[a(1/x_0 - 1/x) + 3b (1/x_0^2 - 1/x^2)\right]  \right\}  - 1}.
\end{equation}
\begin{eqnarray}\label{eq:Ychi-late2}
      Y_{\chi}(x) = 
  \frac{\eta}{ 1 -  \left\{ Y_{\bar\chi}(x_0)/[\eta + Y_{\bar\chi}(x_0)] \right\}\,
  {\rm exp}\left\{ \eta \lambda_1
  \left[a(1/x_0 - 1/x) + 3b (1/x_0^2 - 1/x^2)\right]  \right\} }\,.
\end{eqnarray}

The final relic density is 
\begin{equation}
       \Omega_{\rm DM} h^2 
   \simeq 2.76 \times 10^8\, m \left[ Y_{\chi}(x) + Y_{\bar\chi}(x) \right]\,.
\end{equation}
\section{Constraints on parameter space}
The Planck team derived the Dark Matter relic density as  
\cite{Ade:2015xua},
\begin{eqnarray} \label{eq:pldata}
  \Omega_{\rm DM} h^2 = 0.1199 \pm 0.0022\, .
\end{eqnarray}
We use the measured Dark Matter relic density to find constraints on the
paramete space. For asymmetric Dark Matter, particle and anti--particle
contributions have to be added as 
$\Omega_{\rm DM} = \Omega_{\chi}  + \Omega_{\bar\chi}$.

\begin{figure}[h]
  \begin{center}
    \hspace*{-0.5cm} \includegraphics*[width=8.7cm]{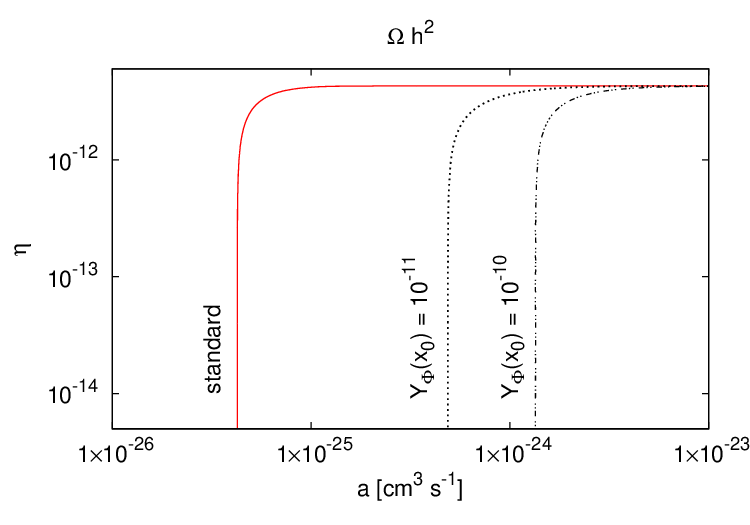}
    \caption{\label{fig:c} 
    \footnotesize Contour plot of $s-$wave annihilation cross section 
    $ a $ and asymmetry factor $\eta$ when 
    $\Omega_{\rm DM} h^2 = 0.1199$. The other parameters are the same as in 
    Fig.\ref{fig:b}.}
     \end{center}
\end{figure}

The contour plot of $s-$wave annihilation cross section and 
asymmetry factor $\eta$ is shown in Fig.\ref{fig:c} when 
$\Omega_{\rm DM} h^2 = 0.1199$. This figure is based on the numerical
solutions of the Bolzmann equations
(\ref{eq:Ychifin}) and (\ref{eq:Ychibarfin}). 
The dotted (black) and dashed (black) curves
are for the case of including the decay of heavier particle to asymmetric 
Dark Matter. The solid (red) curve is for the standard case. We find that the 
required cross section for $Y_{\phi}(x_0) = 10^{-11}$ is nearly one order of
magnitude larger than the standard one in order to fall into the observational 
range. The increased relic density of asymmetric Dark Matter due to the decay 
of heavier particle is suppressed by the larger cross section. Furthermore, 
the figure shows that the needed cross section is slightly larger when the 
initial value of $Y_{\phi}$ is large. We note that the horizontal part of the 
contours is not affected by the decay of heavier particle. The 
anti--particle Dark Matter density of $Y_{\bar\chi}$ is exponentially
suppressed when the asymmetry increases and the total Dark Matter relic
density is determined by the particle density as 
$\Omega_{\rm DM} \sim m Y_{\chi} \sim m \eta$. Therefore, the final relic 
density of DM is independent of the cross section $a$. 


\section{Summary and conclusions}

The relic abundance of asymmetric Dark Matter including the decay of heavier
particles is discussed in low reheating temperature scenario. Here we assume 
there is asymmetry before the unstable heavy particles decay into asymmetric 
Dark Matter particle and anti--particle in the same amount. 
We found that relic abundances of asymmetric Dark Matter particle and 
anti--particle are both increased when there is additional contribution from 
the decay of heavier particles. The increases of particle and 
anti--particle abundances due to the non--thermal production depends on the 
decay rate of heavy particles and slightly on the initial value of heavy
particle abundance. Comparison with the usual thermal production shows that
the depletion of the anti--particle abundance is slower for the case of
including decay of heavy particles. 

In the end, we used the observed DM abundance to obtain the constraints on 
the annihilation cross section and asymmetry factor when there is contribution
from the decay of heavier particles to asymmetric Dark Matter. We found the 
required annihilation cross section with the decay of heavy particles is 
almost one order of magnitude larger than the case of without including 
non--thermal production. Those results are important for asymmetric Dark 
Matter. There is possibility to detect asymmetric Dark Matter in indirect 
detection due to the increased amount of anti--particle relic density when 
there is additional contribution from the decay of heavy particles. 

\section*{Acknowledgments}

The work is supported by the National Natural Science 
Foundation of China (2020640017, 2022D01C52, 11765021).

\end{document}